\documentclass{article}%
\usepackage{spconf,graphicx}
\usepackage{amssymb,amsthm,amsfonts}
\usepackage[cmex10]{amsmath}
\usepackage{bm}

\title{Exact Performance Analysis of the Oracle Receiver for Compressed Sensing Reconstruction}
\name{Giulio Coluccia$^\star$, Aline Roumy$^\dagger$, Enrico Magli$^\star$\thanks{This work has been supported by the European Research Council under the European Community's Seventh Framework Programme (FP7/2007-2013) / ERC Grant agreement n° 279848}}
\address{$^\star$ Politecnico di Torino, Italy\\
		 $^\dagger$ INRIA, France}
\newcommand{\mat}[1]{\ensuremath{\bm{\mathrm{#1}}}}

\newcommand{\I}{\ensuremath{\mat{I}}}
\newcommand{\x}{\ensuremath{\mat{x}}}
\newcommand{\thet}{\ensuremath{\bm{\theta}}}
\newcommand{\z}{\ensuremath{\mat{z}}}

\newcommand{\y}{\ensuremath{\mat{y}}}

\newcommand{\Ps}{\ensuremath{\bm{\Psi}}}

\newcommand{\Ph}{\ensuremath{\mat{\Phi}}}

\newcommand{\0}{\ensuremath{\mat{0}}}
\newcommand{\U}{\ensuremath{\mat{U}}}
\newcommand{\X}{\ensuremath{\mat{X}}}

\newcommand{\UO}{\ensuremath{\mat{U}_{\Omega}}}
\newcommand{\UpO}{\ensuremath{\mat{U}^\dagger_{\Omega}}}
\newcommand{\UpOT}{\ensuremath{\mat{U}^{\dagger \ T}_{\Omega}}}

\newcommand{\R}{\ensuremath{\mathbb{R}}}

\newcommand{\var}{\ensuremath{\sigma^2}}

\newcommand{\trasp}[1]{\ensuremath{#1 ^\mathsf{T}}}

\newcommand{\trace}[1]{\ensuremath{\mathrm{Tr}\left(#1\right)}}
\newcommand{\mean}[1]{\ensuremath{\mathbb{E}\left[#1\right]}}
\newcommand{\meanover}[2]{\ensuremath{\mathbb{E}_{#1}\left[#2\right]}}

\newcommand{\N}{\mathcal{N}}
\newcommand{\W}{\mathcal{W}}

\newcommand{\inv}[1]{\ensuremath{{#1}^{-1}}}

\def\Ri{\mathbb{R}}

\newcommand{\lzeronorm}[1]{\ensuremath{\left\| #1\right\|_{0}}}
\newcommand{\lonenorm}[1]{\ensuremath{\left\| #1\right\|_{1}}}
\newcommand{\ltwonorm}[1]{\ensuremath{\left\| #1\right\|_{2}}}

\newtheorem{lemma}{Lemma}
\newtheorem{theorem}[lemma]{Theorem}

\begin{document}
%
\maketitle
\begin{abstract}

\end{abstract}
A sparse or compressible signal can be recovered from a certain number of noisy random projections, smaller than what dictated by classic Shannon/Nyquist theory. In this paper, we derive the \emph{closed--form} expression of the mean square error performance of the \emph{oracle} receiver, knowing the sparsity pattern of the signal. With respect to existing bounds, our result is \emph{exact} and does not depend on a particular realization of the sensing matrix. Moreover, our result holds irrespective of whether the noise affecting the measurements is white or correlated. Numerical results show a perfect match between equations and simulations, confirming the validity of the result.

\begin{keywords}
Compressed Sensing, Oracle Receiver, Wishart Matrix
\end{keywords}
\section{Introduction}
\label{sec:intro}

Compressed sensing (CS) \cite{donoho2006cs,candes2006nos} has emerged in past years as an efficient technique for sensing a signal with fewer coefficients than dictated by classic Shannon/Nyquist theory. The hypothesis  underlying this approach is that the signal to be sensed 
must have a sparse -- or at least compressible -- representation in a convenient basis. In CS, sensing is performed by taking a number of linear projections of
the signal onto pseudorandom sequences. Therefore, the acquisition presents appealing properties. First, it requires \textit{low encoding complexity}, since no sorting of the sparse signal representation is required. Second,  the choice of the sensing matrix distribution is blind to the source distribution.

Several different techniques can be used to reconstruct a signal from CS measurements. Often, for performance assessment, the ideal \emph{oracle} receiver, \emph{i.e.,} a receiver with perfect knowledge of the signal sparsity support, is considered as a benchmark. But even for this ideal receiver, only upper and lower performance bounds are available. For example, in \cite{eldar2012compressed} a bound depending on a particular realization of the sensing matrix was derived. This bound represents a worst--case scenario since it depends on the maximum norm of the noise vector. An average (over noise) bound was presented in \cite{DBLP:journals/corr/abs-1104-4842} for white noise and in \cite{laska2012regime} for correlated noise. Both bounds depend on the Restricted Isometry Property (RIP) constant of the sensing matrix, a parameter taking different values from realization to realization of the sensing matrix and whose evaluation represents a combinatorial complexity problem. Even if there exist classes of matrix respecting the RIP with a certain constant with high probability, this would give a probabilistic result, restricted to a specific class of sensing matrices. Moreover, note that \cite{laska2012regime} overestimates the reconstruction error giving a result which depends on the maximum eigenvalue of the noise covariance matrix. Other results can be found in \cite{arias2013fundamental} and \cite{candes2013well}.

In this paper, we present the \emph{exact} average performance of the oracle receiver. The average is taken over noise distribution but also over the sensing matrix distribution, and does not depend on the RIP constant of a specific sensing matrix (or family of sensing matrices), but only on system or signal parameters. Using some recent results about Wishart random matrix theory \cite{Cook2011}, we show that the performance depends, apart from system parameters, on the variance of the noise, only, and not on its covariance. Hence, our result can be applied both to systems where measurements are corrupted by either white or correlated noise.

\section{Background}
\label{sec:background}

%
%

\subsection{Compressed Sensing}
\label{sec:background_CS}

In the standard CS framework, introduced in \cite{donoho2006cs,candes2006nos}, the signal $\x\in\Ri^{N\times 1}$, having a $K$--sparse representation in some basis $\Ps\in\Ri^{N\times N}$, \textit{i.e.}: $\x = \Ps \bm{\theta},\quad \lzeronorm{\bm{\theta}} = K,\quad K\ll N$, can be recovered by a smaller vector of noisy linear measurements $\y = \Ph\x+\z$, $\y\in\Ri^{M\times 1}$ and $K<M<N$,  where $\Ph\in\Ri^{M\times N}$ is the \emph{sensing matrix} and $\mat{z}\in\Ri^{M\times 1}$ is the vector representing additive noise such that $\ltwonorm{\mat{z}} < \varepsilon$, by solving the $\ell_1$ minimization with inequality constraints 
\begin{equation}\label{eq:CS_recovery_relaxed}
\widehat{\bm{\theta}}=\arg\min_{\bm{\theta}}\lonenorm{\bm{\theta}}\ \quad \text{s.t.}\quad \ltwonorm{\Ph\Ps\bm{\theta} - \y} < \varepsilon~
\end{equation}
and $\widehat{\x}=\Ps\widehat{\bm{\theta}}$, known as basis pursuit denoising, provided that $M = O(K\log(N/K))$ and that each submatrix consisting of $K$ columns of $\Ph\Ps$  is (almost) distance preserving \cite[Definition 1.3]{eldar2012compressed}. The latter condition is the \emph{Restricted Isometry Property} (RIP). Formally, the matrix $\Ph\Ps$ satisfies the RIP of order $K$ if $\exists \delta_K \in (0,1]$ such that, for any $\bm{\theta}$ with $\lzeronorm{\bm{\theta}} \le K$:
\begin{equation}
(1-\delta_K)\ltwonorm{\bm{\theta}}^2\le\ltwonorm{\Ph\Ps\bm{\theta}}^2\le(1+\delta_K)\ltwonorm{\bm{\theta}}^2,
	\label{eq:RIP}
\end{equation}
where $\delta_K$ is the RIP constant of order $K$. It has been shown in \cite{baraniuk2008spr} that when $\Ph$ is an i.i.d. random matrix drawn from any subgaussian distribution and $\Ps$ is an orthogonal matrix, $\Ph\Ps$ satisfies the RIP with overwhelming probability.

\subsection{Wishart Matrices}
\label{sec:background_wishart}
Let $\x_i$ be a zero--mean Gaussian random vector with covariance matrix $\bm{\Sigma}$. Collect $n$ realizations of $\x_i$ as rows of the $n\times p$ matrix $\X$. Hence, $\trasp{\X}\X$ is distributed as a $p$-dimensional Wishart matrix with scale matrix $\bm{\Sigma}$ and $n$ degrees of freedom \cite{press1982applied}:
$$
\mat{W} = \trasp{\X}\X\sim \W_p\left(\bm{\Sigma},n\right)~.
$$
When $n>p$, $\mat{W}$ can be inverted. The distribution of $\inv{\mat{W}}$ is the \emph{Inverse Wishart}, whose distribution and moments were derived in \cite{von1988moments}:
$$
\inv{\mat{W}} \sim \inv{\W}_p\left(\inv{\bm{\Sigma}},n\right)~.
$$
On the other hand, when $n<p$, $\mat{W}$ is rank--deficient, hence not invertible. Its Moore--Penrose pseudoinverse $\mat{W}^\dagger$ follows a generalized inverse Wishart distribution, whose distribution is given in \cite{DiazGarcia2006} and mean and variance were recently derived in \cite[Theorem 2.1]{Cook2011}, under the assumptions that $p>n+3$ and $\mat{\Sigma} = \mat{I}$.

\section{Performance of the Oracle Receiver}
\label{sec:oracle_performance}

\subsection{System model}
\label{sec:system_model}

Consider the vector $\x = \Ps \bm{\theta} \in \R^N$. The nonzero components of the $K$--sparse vector $\thet$ are modeled as i.i.d. centred random variables with variance  $\var_{\theta}$.

The vector $\x$  is observed through a smaller vector of noisy Gaussian measurements defined as the vector $\y\in\Ri^{M}$ such that $\y = \Ph\x +\z$,  where the sensing matrix $\Ph\in\Ri^{M \times N}$, with $M<N$, is a random matrix with i.i.d. entries drawn from a zero--mean Gaussian distribution with variance $\var_{\Phi}$ and $\z\in\Ri^{M\times 1}$, representing the noise, is drawn from a zero--mean multivariate random distribution with covariance matrix $\mat{\Sigma}_z$.

We remark here that in our analysis we consider measurements affected both by white noise, \emph{i.e.,} the case where $\mat{\Sigma}_z = \mat{I}$, like thermal noise or quantization noise deriving from uniform scalar quantizer in the high--rate regime, as well as correlated noise, like the one affecting measurements quantized using vector quantization or the noise at the output of a low--pass filter.

\subsection{Error affecting the oracle reconstruction}
\label{subsec:RD_CS reconstruction}

We now evaluate the performance of CS reconstruction with noisy  measurements. The performance depends on the amount of noise affecting the measurements. In particular, the distortion ${\ltwonorm{\widehat{\x}-\x}^2}$ is upper bounded by the
noise variance up to a scaling factor \cite{candes2006ssr, candes2008restricted} $\ltwonorm{\widehat{\x}-\x}^2 \le c^2\varepsilon^2$, where the constant $c$  depends on the realization of the measurement matrix, since it is a function of the RIP constant. Since we consider the average\footnote{The average performance is obtained averaging over all random variables \emph{i.e.} the measurement matrix, the non-zero components $\thet$ and noise, as for example in \cite{laska2012regime}.} performance, we need to consider the worst case $c$ and this upper bound will be very loose \cite[Theorem 1.9]{eldar2012compressed}.

Here, we consider the \emph{oracle} estimator, which is the estimator knowing exactly the sparsity support $\Omega=\{i|\bm{\theta}_i\neq 0\}$ of the signal $\x$. 

Let $\mat{U}_{\Omega}$ be the submatrix of $\U$ obtained by keeping the columns of $\Ph\Ps$ indexed by $\Omega$, and let $\Omega^c$ denote the complementary set of indexes. The optimal reconstruction is then obtained by using the pseudo--inverse of $\mat{U}_{\Omega}$, denoted by $\U_{\Omega}^\dagger$:
\begin{align}\label{eq:rec_oracle}
&\left\{\begin{array}{ll}
\widehat\thet_{\Omega} & =  
 \mat{U}^\dagger_{\Omega} \y :=  \left(\trasp{\mat{U}_{\Omega}}\mat{U}_{\Omega}\right)^{-1}\trasp{\mat{U}_{\Omega}}\y \\
\widehat\thet_{\Omega^c} & = \0 
\end{array}\right.\\
&\widehat\x = \Ps \widehat\thet
\end{align}

For the oracle estimator, upper and lower bounds depending on the RIP constant can be found, for example in \cite{DBLP:journals/corr/abs-1104-4842} when the noise affecting the measurements is white and in \cite{laska2012regime} when the noise is correlated. Unlike \cite{DBLP:journals/corr/abs-1104-4842, laska2012regime}, in this paper the average performance of the oracle, depending on system parameters only, is derived exactly. Relations with previous work will be thoroughly described in section~\ref{sec:prev_work}.

As we will show in the following sections, the characterization of the ideal oracle estimator allows to derive the reconstruction RD functions with results holding also when non ideal estimators are used.

\begin{theorem} 
\label{th:RD reconstruction non distributed}
Let $\x$ and $\y$ be defined as in section~\ref{sec:system_model}. Assume reconstruction by the oracle estimator, when the support $\Omega$ of $\x$ is available at the receiver. The average reconstruction error of any reconstruction algorithm is lower bounded by that of the oracle estimator that satisfies
\begin{equation}\label{eq:theo_1}
\mean{\ltwonorm{\widehat\x-\x}^2} = \frac{K}{M (M-K-1)} \frac{\trace{\mat{\Sigma}_z}}{\sigma^2_\Phi}
\end{equation}
\end{theorem}

\noindent \textbf{{Proof.}} We derive a lower bound on the achievable distortion by assuming that the sparsity support $\Omega$ of $\x$ is known at the decoder. 

Hence, 
\begin{align}
\mean{\ltwonorm{\widehat\x-\x}^2} &= \mean{\ltwonorm{\widehat\thet-\thet}^2} =  \mean{\ltwonorm{\widehat\thet_{\Omega}-\thet_{\Omega}}^2} 
\label{eq:rc 1}\\
&= \mean{\ltwonorm{ \UpO \z}^2}
\label{eq:rc 2}\\
&= \mean{\z^T \mean{(\UO\UO^T)^\dagger} \z }  
\label{eq:rc 3}
\end{align}
The first equality in \eqref{eq:rc 1} follows from the orthogonality of the matrix $\Ps$, whereas the second one follows from the assumption that $\Omega$ is the true support of $\thet$. \eqref{eq:rc 2} comes from the definition of the pseudo-inverse, and \eqref{eq:rc 3} from the equality $\UpOT \UpO = (\UO\UO^T)^\dagger$ and from the statistical independence of \U\ and \z. Then, if $M>K+3$,
\begin{align}
\mean{\ltwonorm{\widehat\x-\x}^2} &= \mean{\z^T \frac{K}{M (M-K-1)} \frac{1}{\sigma^2_\Phi} \I \  \z }
\label{eq:rc 4}\\
&= \frac{K}{M (M-K-1)} \frac{\trace{\mat{\Sigma}_z}}{\sigma^2_\Phi}
\end{align}
where \eqref{eq:rc 4} comes from the fact that, since $M>K$, $\UO \UO^T$ is rank deficient and follows a singular $M$-variate Wishart distribution with $K$ degrees of freedom and scale matrix $\sigma^2_\Phi \I$ \cite{DiazGarcia2006}. Its pseudo-inverse follows a generalized inverse Wishart distribution, whose distribution is given in \cite{DiazGarcia2006} and the mean value is given in \cite[Theorem 2.1]{Cook2011}, under the assumption that $M>K+3$. Note that the condition $M>K+3$ is not restrictive since it holds for all $K$ and $M$ of practical interest. It can be noticed that the distortion of the oracle only depends on the variance of the elements of \z\ and not on its covariance matrix. Therefore, our result holds even if the noise is correlated (for instance if vector quantization is used).  As a consequence, we can apply our result to any quantization algorithm or to noise not resulting from quantization. Note that, if the elements of \z\ have the same variance, \eqref{eq:theo_1} reduces to
\begin{equation}\label{eq:theo_1_1}
\mean{\ltwonorm{\widehat\x-\x}^2} = \frac{K}{M-K-1}  \frac{\sigma_z^2}{\sigma^2_\Phi}
\end{equation}
 \hfill $\square$

\subsubsection{Relations with previous work}
\label{sec:prev_work}

The results obtained in Theorem~\ref{th:RD reconstruction non distributed} provide a twofold contribution with respect to results already existing in literature about the oracle reconstruction. First, they are exact and not given as bounds. Second, they do not depend on parameters which cannot be evaluated in practical systems, \emph{e.g.,} the RIP constant of the sensing matrices. For example, in \cite{eldar2012compressed} the following worst--case upper bound was derived
\begin{equation}
\ltwonorm{\widehat{\x}-\x}^2 \le \frac{1}{1-\delta_{2K}}\ltwonorm{\z}^2~,
\end{equation}
which depends on a particular realization of the sensing matrix, since it depends on its RIP constant $\delta_{2K}$, and is very conservative, since it is function of the maximum $\ell_2$ norm of the noise vector. An average evaluation (over noise) was given in \cite[Theorem 4.1]{DBLP:journals/corr/abs-1104-4842} where the performance of the oracle receiver with measurements affected by white noise was derived
\begin{equation}\label{eq:oracle_liter}
\frac{K}{1+\delta_{K}}\var_{z}\le \meanover{\z}{\ltwonorm{\widehat{\x}-\x}^2} \le \frac{K}{1-\delta_{K}}\var_{z}
\end{equation}
but still the equation depends on the RIP constant of the sensing matrix and hence, on a particular realization. The result of \eqref{eq:oracle_liter} was generalized in \cite{laska2012regime} to correlated noise
\begin{equation}\label{eq:oracle_corr_liter}
\meanover{\z}{\ltwonorm{\widehat{\x}-\x}^2} \le \frac{K}{1-\delta_{K}}\lambda_{\max}(\mat{\Sigma}_z)~,
\end{equation}
where $\mat{\Sigma}_z$ is the covariance matrix of \z\ and $\lambda_{\max}(\cdot)$ represents the maximum eigenvalue of the argument. Hence, \eqref{eq:oracle_corr_liter} represents an even looser bound, since the contribution of the noise correlation is upper bounded by using its biggest eigenvalue.

Finally, the results of Theorem~\ref{th:RD reconstruction non distributed} can help to generalize related results, \emph{e.g.,} the Rate--Distortion performance of systems based on Compressed Sensing. See for example \cite[section III.C]{dai2011quantized}, where a lower bound is derived, or \cite{coluccia2013operational}, where the exact RD performance is derived. 

\section{Numerical Results}
\label{sec:num_res}

In this section, we show the validity of the results of Theorem~\ref{th:RD reconstruction non distributed} by comparing the equations to the results of simulations. Here and in the following sections, signal length is $N=512$ with sparsity $K=16$. $M=128$ measurements are taken. The nonzero elements of the signal are distributed as $\N(0,1)$. The sparsity basis \Ps\ is the DCT matrix. The sensing matrix is composed by i.i.d. elements distributed as zero--mean Gaussian with variance $1/M$. The noise vector is Gaussian with zero mean, while the covariance matrix depends on the specific test and will be discussed later. The reconstructed signal $\widehat{\x}$ is obtained using the oracle estimator. A different realization of the signal, noise and sensing matrix is drawn for each trial, and the reconstruction error, evaluated as $\mean{\ltwonorm{\widehat\x-\x}^2}$, is averaged over 1,000 trials.

\subsection{White noise}

\begin{figure}
\centering
\vspace*{-5mm}
\includegraphics[width=0.95\columnwidth]{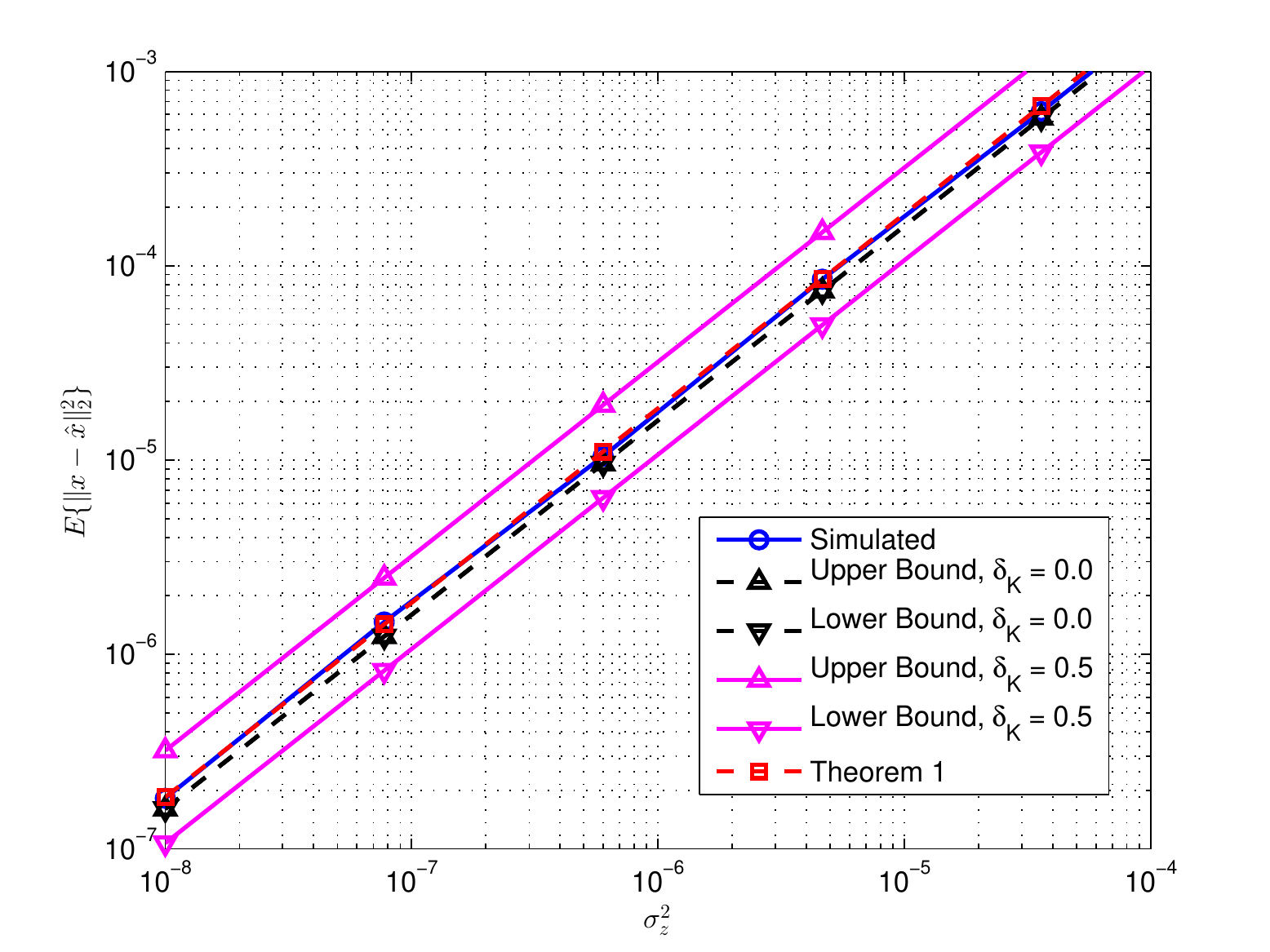}
\vspace*{-5mm}\caption{\small{Oracle reconstruction error. Simulations vs. Theorem~\ref{th:RD reconstruction non distributed}. $N=512$, $K=16$, $M=128$. White noise: $\mat{\Sigma}_z = \var_z\mat{I}_M$.}}
\label{fig:oracle_white}
\end{figure}

In this first experiment, the measurement vector \y\ is corrupted by white Gaussian noise, \emph{i.e.,} $\z\sim\N_p(\mat{0}, \var_z\mat{I}_M)$. Fig.~\ref{fig:oracle_white} shows the comparison between the simulated reconstruction error and \eqref{eq:theo_1_1}. It can be easily noticed that the match between simulated and theoretical curve is perfect. As a term of comparison, we plot also the upper and lower bounds of \eqref{eq:oracle_liter}, for $\delta_K = 0$ (ideal case) and $\delta_K = 0.5$. It can be noticed that for $\delta_K = 0$ the two bounds match and are close to the simulated curve, but even the upper bound is lower than the real curve. Instead, for $\delta_K = 0.5$ the two bounds are almost symmetric with respect to the realistic curve but quite far from it. The conclusion is that bounds in the form of \eqref{eq:oracle_liter} are difficult to use due to the lack of knowledge of the RIP constant. Even if the sensing matrix belongs to a class where a probabilistic expression of the RIP constant exists, like the ones in \cite{vershynin2012nonasymptotic}, a specific value depending on system parameters only is usually difficult to obtain since it depends on constants whose value is unknown or hard to compute. Tests with generic diagonal $\mat{\Sigma}_z$ have also been run, confirming a perfect match with \eqref{eq:theo_1}.

\subsubsection{Uniform scalar quantization}

\begin{figure}
\centering
\vspace*{-2mm}
\includegraphics[width=0.95\columnwidth]{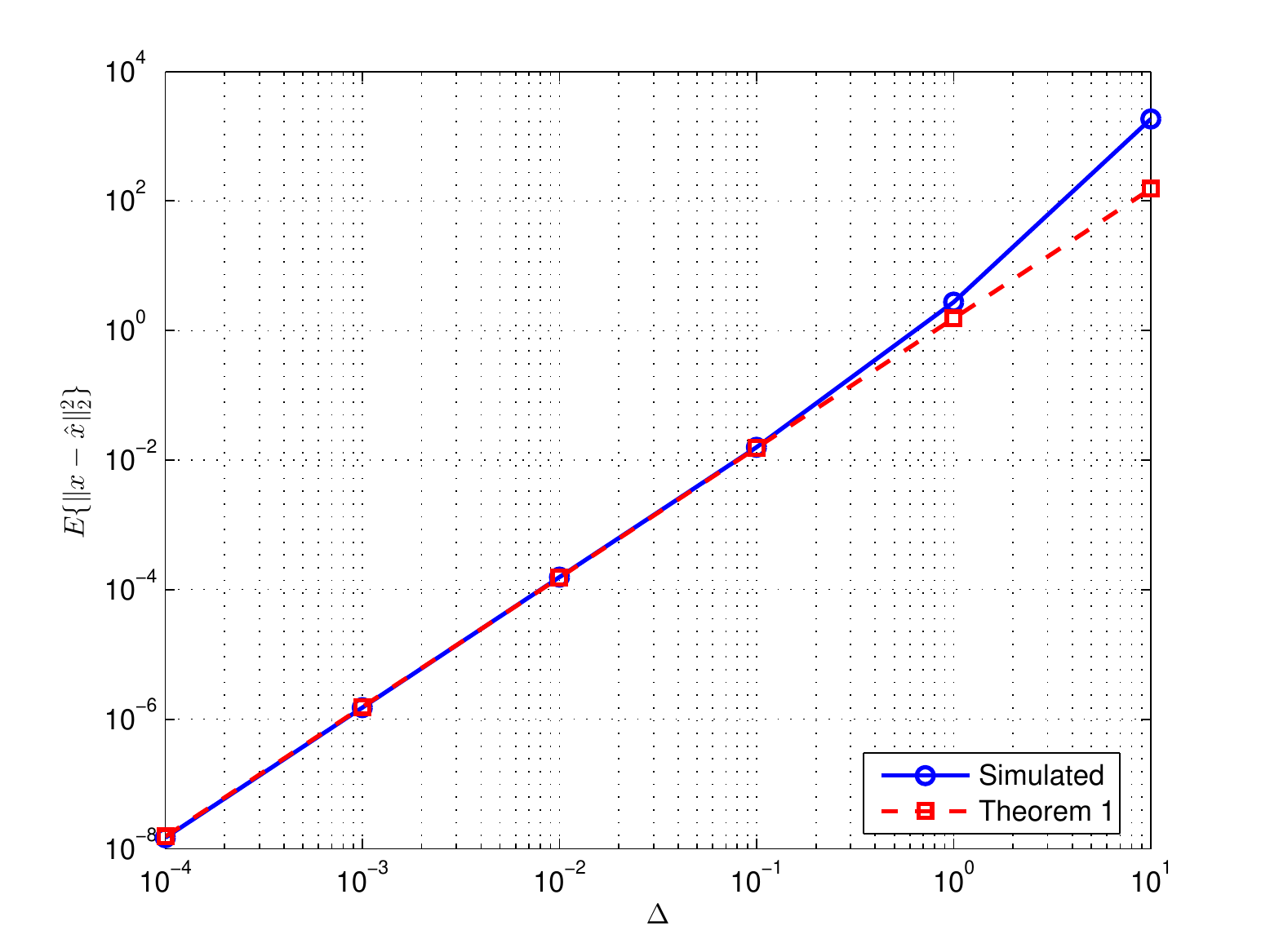}
\vspace*{-5mm}\caption{\small{Oracle reconstruction error. Simulations vs. Theorem~\ref{th:RD reconstruction non distributed}. $N=512$, $K=16$, $M=128$. Measurements quantized with Uniform Scalar Quantizer with step size $\Delta$.}}
\label{fig:oracle_white_quant}
\end{figure}

A practical application of the white noise case is a system where the measurement vector is quantized using an uniform scalar quantizer with step size $\Delta$. In this case, equation \eqref{eq:theo_1_1} is very handy because it is well known that in the high--rate regime the quantization noise can be considered as uncorrelated and its variance is equal to $\frac{\Delta^2}{12}$. In Fig.~\ref{fig:oracle_white_quant}, we plot the reconstruction error of the oracle from quantized measurements vs. the step size $\Delta$. It can be noticed that the match between simulations and proposed equation is perfect in the high--rate regime, \emph{i.e.,} when the step size gets small.

\subsection{Correlated noise}

\begin{figure}
\centering
\vspace*{-5mm}
\includegraphics[width=0.95\columnwidth]{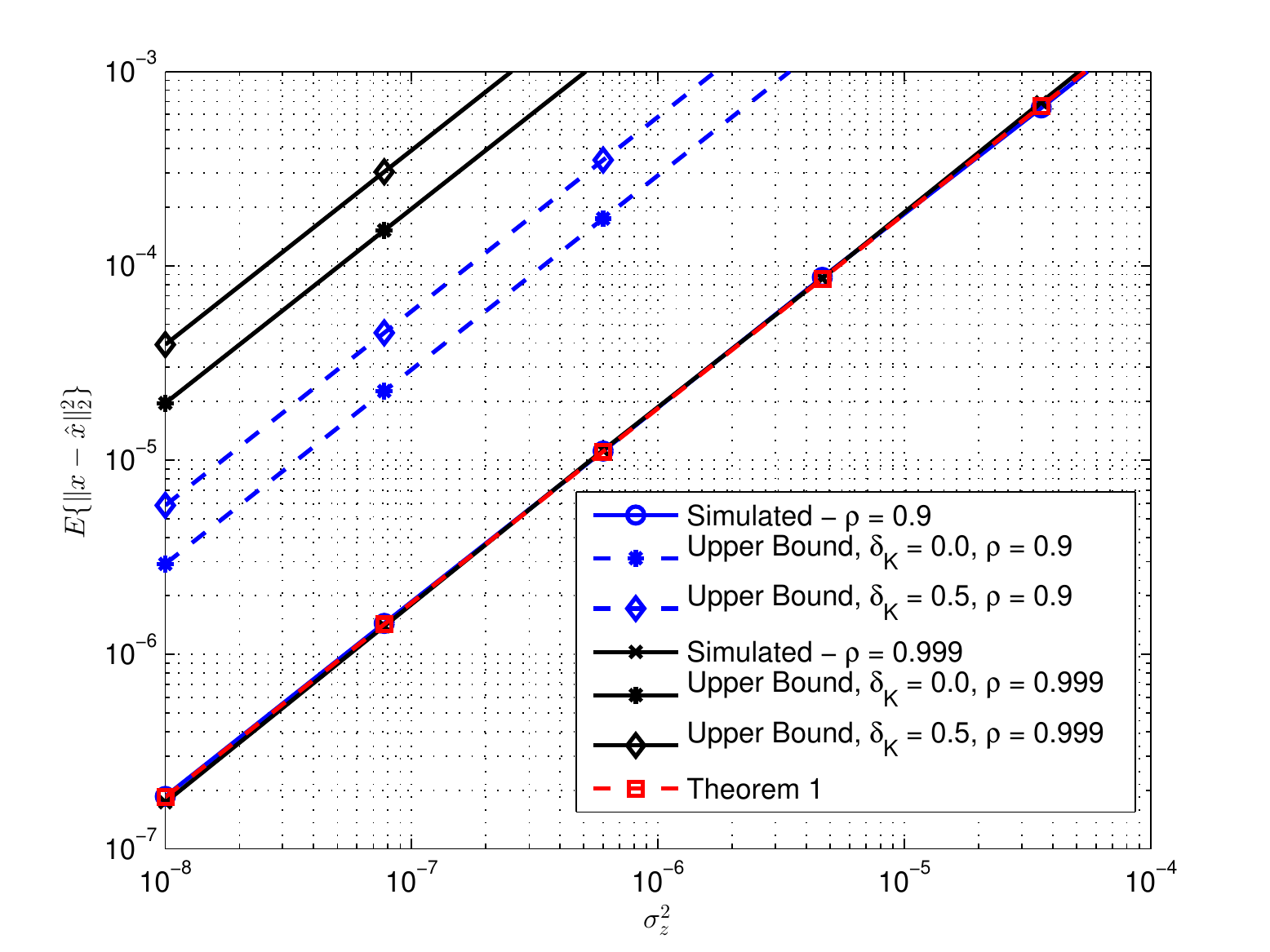}
\vspace*{-5mm}\caption{\small{Oracle reconstruction error. Simulations vs. Theorem~\ref{th:RD reconstruction non distributed}. $N=512$, $K=16$, $M=128$. Correlated noise: ($\mat{\Sigma}_z)_{i,j} = \var_z\rho^{|i-j|}$ and $\rho = 0.9,\ 0.999$.}}
\label{fig:oracle_corr}
\end{figure}

We also report in Fig.~\ref{fig:oracle_corr} the results obtained reconstructing with the oracle receiver the measurements corrupted by correlated noise. In particular, the $i,j$-th element of the noise covariance matrix will be given by $(\mat{\Sigma}_z)_{i,j} = \var_z\rho^{|i-j|}$. The correlation coefficient takes the values of $\rho = 0.9$ and $0.999$. We compare the simulations with \eqref{eq:theo_1} and with the upper bound of \eqref{eq:oracle_corr_liter}, for $\delta_K = 0$ (ideal case) and $\delta_K = 0.5$. First, it can be noticed from Fig.~\ref{fig:oracle_corr} that simulations confirm the result that the performance of the oracle does not depend on noise covariance but only on its variance. This is shown by the fact that simulations for $\rho=0.9$ overlap the ones for $\rho = 0.999$, and both match \eqref{eq:theo_1}, confirming the validity of Theorem~\ref{th:RD reconstruction non distributed} even in the correlated noise scenario. Second, Fig.~\ref{fig:oracle_corr} shows that the upper bounds of \eqref{eq:oracle_corr_liter} highly overestimate the real reconstruction error of the oracle, even for the ideal $\delta_K = 0$ case. This can be explained by considering that in \eqref{eq:oracle_corr_liter}, for the chosen correlation model, $\lambda_{\max}$ tends to $\var_zM$ when $ \rho$ tends to $1$. 

\section{Conclusions and future work}
\label{sec:conclusions}

In this paper, we derived the closed--form expression of the average performance of the \emph{oracle receiver} for Compressed Sensing. Remarkably, this result is exact, and does not depend on the RIP constant or the noise covariance matrix. We showed that the theoretical results perfectly match the ones obtained by numerical simulations. This represents a significant improvement with respect to existing results, which consist in bounds depending on parameters that are hardly available.

As a future activity, this work can be extended to non ideal receivers, with a mismatched knowledge of the signal sparsity pattern. In that case, the performance will depend both on the noise affecting the signal and on the number of misestimated position in the sparsity pattern.


\clearpage

\bibliographystyle{IEEEbib}
\bibliography{discos_spt}

\end{document}